\documentclass[journal]{IEEEtran}

\usepackage{graphicx}
\usepackage{booktabs}
\usepackage{subfigure}
\usepackage{overpic}
\usepackage{amsmath}
\usepackage{flushend}
\usepackage{amssymb}
\usepackage{multirow}
\usepackage{makecell}
\usepackage{color, soul}

\soulregister\cite7

\usepackage[ruled]{algorithm2e}
\usepackage{hyperref}
\hypersetup{
	colorlinks=true,
	linkcolor=blue,
	urlcolor=blue,
	citecolor=blue,
}

\usepackage{colortbl}
\usepackage[dvipsnames]{xcolor}
\definecolor{Bg}{HTML}{e0f1ff}

\begin{document}

\title{Wavelet-based Bi-dimensional Aggregation Network for SAR Image Change Detection}
\author{Jiangwei Xie, Feng Gao, Xiaowei Zhou, Junyu Dong
\thanks{This work was supported in part by the National Science and Technology Major Project under Grant 2022ZD0117202, in part by the Natural Science Foundation of Qingdao under Grant 23-2-1-222-ZYYD-JCH, and in part by the Postdoctoral Fellowship Program of CPSF under Grant GZC20241614. (\textit{Corresponding author: Xiaowei Zhou.})

Jiangwei Xie, Feng Gao, Xiaowei Zhou, and Junyu Dong are with the School of Computer Science and Technology, Ocean University of China, Qingdao 266100, China.}}

\markboth{IEEE GEOSCIENCE AND REMOTE SENSING LETTERS}
{Shell}

\maketitle

\begin{abstract}

Synthetic aperture radar (SAR) image change detection is critical in remote sensing image analysis. Recently, the attention mechanism has been widely used in change detection tasks. However, existing attention mechanisms often employ down-sampling operations such as average pooling on the Key and Value components to enhance computational efficiency. These irreversible operations result in the loss of high-frequency components and other important information. To address this limitation, we develop Wavelet-based Bi-dimensional Aggregation Network (WBANet) for SAR image change detection. We design a wavelet-based self-attention block that includes discrete wavelet transform and inverse discrete wavelet transform operations on Key and Value components. Hence, the feature undergoes downsampling without any loss of information, while simultaneously enhancing local contextual awareness through an expanded receptive field. Additionally, we have incorporated a bi-dimensional aggregation module that boosts the non-linear representation capability by merging spatial and channel information via broadcast mechanism. Experimental results on three SAR datasets demonstrate that our WBANet significantly outperforms contemporary state-of-the-art methods. Specifically, our WBANet achieves 98.33\%, 96.65\%, and 96.62\% of percentage of correct classification (PCC) on the respective datasets, highlighting its superior performance. Source codes are available at \url{https://github.com/summitgao/WBANet}.

\end{abstract}


\IEEEpeerreviewmaketitle

\section{Introduction}

\IEEEPARstart{S}{ynthetic} aperture radar (SAR) is adept at producing high-resolution images of the Earth's surface, even under conditions of low visibility caused by adverse weather \cite{wang22graph}. SAR sensors can penetrate cloud cover, making them especially valuable for Earth observation in cloudy or rainy areas. Consequently, SAR data has garnered significant interest from the research community, supporting a range of applications such as object detection \cite{Sevo2016grsl}, disaster assessment \cite{Sarkar2023tgrs}, change detection \cite{yan23tgrs}, and image classification \cite{Qian2021tgrs}. Among these, change detection serves as a crucial tool for identifying changes in land cover, urban growth, and deforestation.

Recently, various convolutional neural network based models for change detection have been developed, demonstrating significant advancements in performance. Hou et al. \cite{hou2020tgrs} introduced an end-to-end dual branch architecture that merges CNN with a generative adversarial network (GAN), enhancing the detection of fine-grained changes. Wang et al. \cite{wang2022jstar} introduced distinctive patch convolution combined with random label propagation, achieving high accuracy in change detection at a reduced computational cost. Zhao et al. \cite{Zhao2023grsl} utilized a multidomain fusion module that integrates spatial and frequency domain features into complementary feature representations. Zhu et al. \cite{Zhu2023grsl} designed a feature comparison module that limits the number of feature channels in the fusion process, enabling better utilization of fine-grained information in the multiscale feature map for more accurate prediction.

The previously mentioned CNN-based methods have demonstrated impressive achievements. Furthermore, Vision Transformer (ViT) \cite{vit} has showcased high performance in various computer vision tasks, leading to the adoption of attention mechanisms in change detection models. Zhang et al. \cite{CAMixer} combine convolution and attention mechanisms to improve the performance of SAR image change detection. Although these pioneer efforts have achieved promising performance, designing an attention-based network for SAR change detection is still a non-trivial task, due to the following reasons: \textit{1) High-frequency information loss in self-attention computation.} Traditional down-sampling methods on the Key and Value components in efficient attention mechanisms, often result in the loss of high-frequency components like texture details. \textit{2) Limitation in non-linear feature transformation.} Existing methods require MLP-like structure for non-linear feature transformation. However, spatial and channel-wise attentions are rarely exploited simultaneously. 

To address the above two limitations, we propose a \textbf{W}avelet-based \textbf{B}i-dimensional \textbf{A}ggregation \textbf{Net}work, WBANet for short, which achieves down-sampling without information dropping and fuses both spatial and channel information. Specifically, we design a Wavelet-based Self-attention Module (WSM) which uses Discrete Wavelet Transform (DWT) and Inverse Discrete Wavelet Transform (IDWT) to enable lossless and invertible down-sampling in the self-attention computation. In addition, we develop a Bi-dimensional Aggregation Module (BAM) to enhance the non-linear feature representation capabilities. This module efficiently captures both spatial and channel-wise feature dependencies.

\begin{figure*}[htbp]
\centering
\includegraphics[width=6.7in]{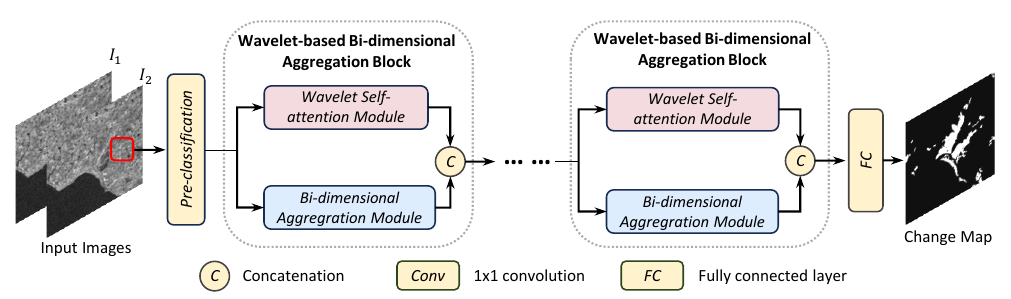}
\caption{An overview of the proposed Wavelet-based Bi-dimensional Aggregation Network (WBANet). The WBANet comprises pre-classification module, wavelet-based bi-dimensional aggregation blocks. Each wavelet-based bi-dimensional aggregation block has two critical components: Wavelet-based Self-attention Module and Bi-dimensional Aggregation Module.}
\label{fig:frame}
\end{figure*}

In summary, the contributions of this letter can be summarized as follows:

\begin{itemize}

\item We propose WSM to integrates DWT and IDWT for down-sampling without information loss, thus preserving textures and other high-frequency details.
\item We develop BAM that captures both spatial and channel-wise feature dependencies effectively. This module merges information from two branches and enhances the non-linear feature representation capabilities.
\item Extensive experiments are conducted on three public SAR datasets,  demonstrating the efficacy of our proposed WBANet. We have made our code publicly available to benefit other researchers.
\end{itemize}

\section{Methodology}

The framework of the proposed WBANet is illustrated in Fig. \ref{fig:frame}. First of all, two multitemporal SAR images ($I_1$ and $I_2$), captured at different times over the same geographic region, are fed into the network. The objective of the change detection task is to generate a change map, marking changed pixels as "1" and unchanged pixels as "0".  Initially, the pre-classification module uses a logarithmic ratio operator to compute a difference image for pseudo-label generation. Subsequently, the hierarchical fuzzy \textit{c}-means algorithm \cite{gao2016grsl} \cite{lihengchao} is employed to classify pixels into changed, unchanged, and intermediate categories.  Then, some wavelet-based bi-dimensional aggregation block process these data from the pre-classification module. Finally, the output features from this block are passed through fully connected layer to generate the change map.

The wavelet-based bi-dimensional aggregation block is comprised of two components: the Wavelet-based Self-attention Block (WSM) and the Bi-dimensional Aggregation Module (BAM). We will present the details of both modules in the following subsections.

\subsection{Wavelet-based Self-attention Module (WSM)}

\begin{figure}[]
    \centering
    \includegraphics[width=3.6in]{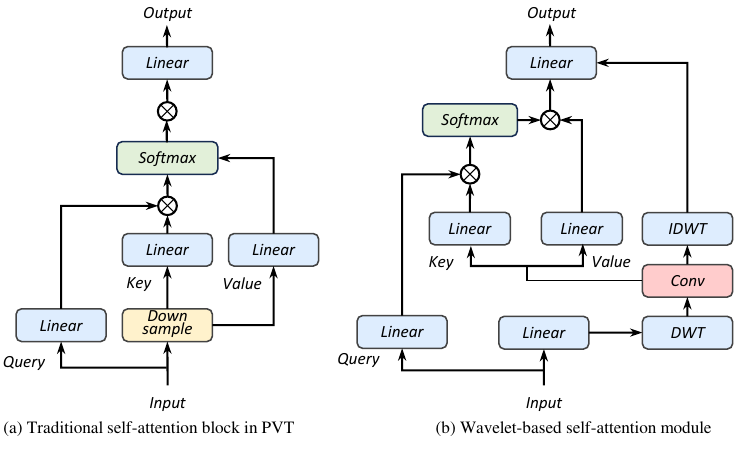}
    \caption{Comparison of the traditional self-attention block in PVT(Pyramid Vision Transformer) and the proposed wavelet-based self-attention module. Linear denotes fully connected layers and $\otimes$ denotes matrix multiplication.}
    \label{fig:waveblock}
\end{figure}

The proposed WSM employs Discrete Wavelet Transform (DWT) and Inverse Discrete Wavelet Transform (IDWT) to facilitate down-sampling in the attention mechanism. The wavelet transform enables feature extraction at both coarse and fine-grained scales, while also ensuring that the down-sampling is invertible. Due to the simple structure and high computational efficiency  of Haar wavelet, it can quickly complete the downsampling operation. Furthermore, SAR change images often exhibit considerable sharp high-frequency information while Haar wavelet is adept at effectively capturing these high-frequency components\cite{mallat1999wavelet}. Thus, we select Haar wavelet to conduct the downsample operation. The structure of this module is shown in Fig. \ref{fig:waveblock}.

To efficiently process the input feature $X \in \mathbb{R}^{H \times W \times C}$, we first reduce its channel dimensions to $\widetilde{X} \in \mathbb{R}^{H \times W \times \frac{C}{4}}$ using a learnable transformation matrix $W_d \in \mathbb{R}^{C \times \frac{C}{4}}$. Following this channel reduction, we apply DWT with the Haar wavelet to down-sample $\widetilde{X}$, and decompose it into four distinct subbands.

Haar wavelet is composed of the low-pass filter $f_L = \left(\frac{1}{\sqrt{2}}, \frac{1}{\sqrt{2}}\right)$ and high-pass filter $f_H = \left(\frac{1}{\sqrt{2}}, -\frac{1}{\sqrt{2}}\right)$. We first encode $\widetilde{X}$ into two subbands $X_L$ and $X_H$ along the rows. Subsequently, these subbands are processed using the same filters along the columns, resulting in four wavelet subbands: $X_{LL}, X_{LH}, X_{HL},$ and $X_{HH}$. Here, $X_{LL} \in \mathbb{R}^{\frac{H}{2} \times \frac{W}{2} \times \frac{C}{4}}$ encodes the low-frequency components, and contains coarse-grained structural information. $X_{LH}, X_{HL}, X_{HH} \in \mathbb{R}^{\frac{H}{2} \times \frac{W}{2} \times \frac{C}{4}}$ represent the high-frequency components, and describe fine-grained textures.

We then concatenate these four subbands along the channel dimension to get $\hat{X}$:
\begin{equation}
\hat{X} = \textrm{Concat}(X_{LL}, X_{LH}, X_{HL}, X_{HH})
\end{equation}

The concatenated output $\hat{X}$ is transformed into Key ($K^w$) and Value ($V^w$) matrices through the convolutional layer, while the Query ($Q$) remains the original input image $X$. In this case, wavelet-based multi-head self-attention computes the interaction across these elements for each head as follows:
\begin{equation}
    \begin{aligned} \text {head}_{i} & = \textrm{Attention}\left({Q}_{i}, {K}_{i}^{{w}}, {V}_{{i}}^{{w}}\right) \\ & =\textrm{Softmax}\left(\frac{{Q}_{i} {K}_{i}^{{w}}}{\sqrt{D_{h}}}\right) {V}_{i}^{{w}}\end{aligned}
\end{equation}
where ${K}_{i}^{w}$ denotes the down-sampled key, ${V}_{i}^{w}$ denotes the down-sampled value, and ${D_{h}}$ represents the dimension of each head.

To enhance the output of the wavelet-based self-attention block, we apply the IDWT to $\hat{X}$ to produce $X^{\boldsymbol{r}}$. The reconstructed $X^{\boldsymbol{r}}$ mirrors the details of the original input image, providing excellent local contextualization and an expanded receptive field. The final output integrates the contributions of each attention head with this reconstructed map. This integration is essential for effectively capturing information across multiple scales.

\begin{figure}[tbp]
    \centering
    \includegraphics[width=3.5in]{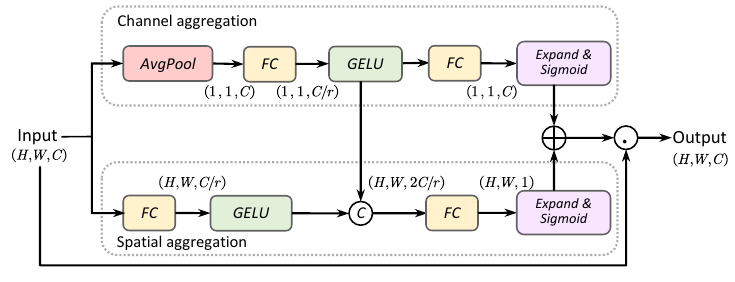}
    \caption{Illustration of the Bi-dimensional Aggregation Module. It includes two branches: global channel attention and local spatial attention. \textcircled{\it{c}} indicates to concatenate in channel axis. $\oplus$ implies element-wise addition and $\odot$ denotes element-wise multiplication.}
    \label{fig:BiAttention}
\end{figure}

The overall operation can be formulated as follows:
\begin{equation}
\textrm{WaveAttn}(X) 
= \textrm{Concat}\left(\textit{head}_{0}, \cdots, \textit{head}_{N_h}, {X}^{r}\right) {W}^{O},
\end{equation}
where ${N_h}$ represents the number of attention heads, and ${W}^{O}$ is the transformation matrix that combines all the heads and the reconstructed image into a single output tensor. The use of wavelet transform in the self-attention mechanism significantly enhances the ability to contextualize information over longer ranges with a reduced computational load compared to conventional self-attention modules. This approach ensures that both global coherence and local detail are preserved and emphasized in the model's outputs.

\subsection{Bi-dimensional Aggregation Module (BAM)}

To enhance the non-linear representation capabilities and effectively capture both spatial and channel dependencies, we develop Bi-dimensional Aggregation Module (BAM), as depicted in Fig. \ref{fig:BiAttention}. This module includes two branches: the channel aggregation branch and the spatial aggregation branch.

\textbf{Channel Aggregation:} In this branch, average pooling is applied in the spatial dimension of input features $X\in\mathbb{R}^{{H\times{W}\times{C}}}$ to aggregate global representations. Subsequently, a fully connected layer (FC) coupled with a GELU activation function reduces the channel dimensions from $C$ to $\frac{C}{r}$, producing an intermediate output $\hat{X}$. Here, $r$, the reduction ratio, is set to 2. This is followed by FC layer and Sigmoid activation function to generate the output of the channel attention branch, $X^{C} \in \mathbb{R}^{1 \times 1 \times C}$.

\textbf{Spatial Aggregation:} First, a linear transformation, together with a GELU activation, transforms the channel dimension to $\frac{C}{r}$, while the spatial dimensions remain unchanged.  The resulting intermediate output, $\widetilde{X}$, is then concatenated with $\hat{X}$ to form $\widetilde{X}^{\prime} \in \mathbb{R}^{H \times W \times \frac{2C}{r}}$. The process culminates similarly to the channel attention branch, resulting in the final output $X^{S} \in \mathbb{R}^{H \times W \times 1}$.

The outputs of both branches, $X^{C}$ and $X^{S}$, are merged through an element-wise summation, ensuring the final output retains the same dimensions as the original input $X$. This integration optimally combines the refined channel and spatial information, enhancing the overall feature representation while maintaining focus on both global context and local details.

\section{Experimental Results and Analysis}

\begin{figure*}[htbp]
  \centering
  \includegraphics[width=7in]{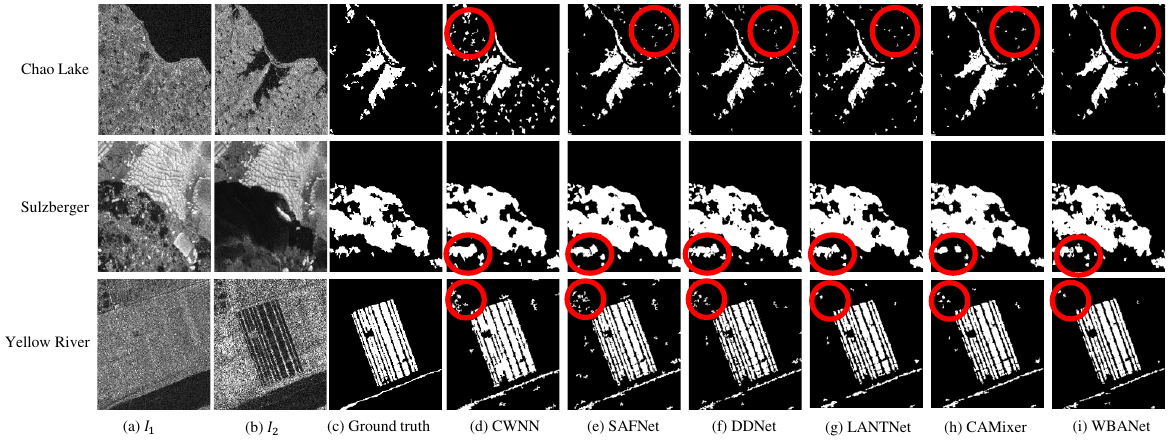}
  \caption{Visualized results of different change detection methods on the three datasets: (a) Image captured at $t_1$. (b) Image captured at $t_2$. (c) Ground truth image. (d)-(i) Results by different methods.}
  \label{fig_result}
\end{figure*}

\subsection{Datasets and Evaluation Metrics}

To validate the effectiveness of the proposed WBANet, we conducted comprehensive experiments on three distinct SAR datasets: Chao Lake, the Yellow River, and Sulzberger Datasets. \textbf{Chao Lake Dataset:} This dataset includes images of Chao Lake in China, captured in May 2020 using the Sentinel-1 sensor. This period coincides with the highest recorded water levels in the lake's history, providing a dynamic range of changes to detect.
\textbf{Sulzberger Dataset:} Captured by the European Space Agency's Envisat satellite over five days in March 2011, this dataset documents the breakup of an ice shelf, offering a unique perspective on drastic natural events. 
\textbf{Yellow River Dataset:} This dataset focuses on the Yellow River Estuary in China, with data collected from June 2008 to June 2009 using the Radarset-2 SAR sensor. This dataset is particularly challenging due to the pronounced speckle noise.

The hierarchical fuzzy c-means algorithm could classify pixels into changed, unchanged, and intermediate categories. Pixels from the changed and unchanged groups are randomly selected as training data, while intermediate group pixels are the test data. For a thorough assessment of our model, we employed five commonly used evaluation metrics: False Positives (FP), False Negatives (FN), Overall Error (OE), Percentage of Correct Classification (PCC), and the Kappa Coefficient (KC). 

\begin{table}[h]
    \centering
	\caption{Change detection results of different methods on three Datasets. The best results are marked in Bold.}
	\label{table_res}
    \begin{tabular}{c|c c c c c} 
    \hline\toprule
    \multirow{2}{*}{Method} & \multicolumn{5}{c}{Results on the Chao Lake dataset} \\
    \cmidrule{2-6}
      & FP & FN & OE & PCC ($\%$)  & KC ($\%$)\\ 
    \midrule
     CWNN \cite{cwnngao} & 7528  & 2213 & 9741 & 93.39 & 65.01\\
     SAFNet \cite{SAFNet} & 2231 & 1272 & 3503 & 97.62 & 85.55\\ 
     DDNet \cite{DDNet} & 1472 & 1559 & 3031 & 97.94 & 87.04\\
     LANTNet \cite{LANTNet} & 1822 & 1023 & 2845 & 98.07 & 88.20\\
     CAMixer \cite{CAMixer} & 1867 & \textbf{906} & 2773 & 98.12 & 88.56\\
    \rowcolor{Bg} Proposed WBANet & \textbf{1092} & 1373 & \textbf{2465} & \textbf{98.33} & \textbf{89.38}\\
    \bottomrule\hline
    \multicolumn{6}{c}{}  \\
    \hline\toprule
    \multirow{2}{*}{Method} 
    & \multicolumn{5}{c}{Results on the Sulzberger dataset}\\
    \cmidrule{2-6}
      & FP & FN & OE & PCC ($\%$)  & KC ($\%$)\\ 
    \midrule
     CWNN \cite{cwnngao} & 2987  & \textbf{387} & 3374 & 94.85 & 86.95\\
     SAFNet \cite{SAFNet} & 1661 & 883 & 2544 & 96.12 & 89.80\\ 
     DDNet \cite{DDNet} & 1835 & 585 & 2420 & 96.31 & 90.39\\
     LANTNet \cite{LANTNet} & 1761 & 635 & 2396 & 96.34 & 90.46\\
     CAMixer \cite{CAMixer} & \textbf{1105} & 1207 & 2312 & 96.47 & 90.56\\
     \rowcolor{Bg} Proposed WBANet & 1553 & 640 & \textbf{2193} & \textbf{96.65} & \textbf{91.23}\\
    \bottomrule\hline
    \multicolumn{6}{c}{}  \\
    \hline\toprule
    \multirow{2}{*}{Method} 
    & \multicolumn{5}{c}{Results on the Yellow River dataset} \\ 
    \cmidrule{2-6}
      & FP & FN & OE & PCC ($\%$)  & KC ($\%$)\\ 
    \midrule
     CWNN \cite{cwnngao} & 3052  & \textbf{1034} & 4086 & 94.50 & 82.46\\
     SAFNet \cite{SAFNet} & 2199 & 1467 & 3666 & 95.06 & 83.69\\ 
     DDNet \cite{DDNet} & 1251 & 2222 & 3473 & 95.32 & 83.76\\
     LANTNet \cite{LANTNet} & 915 & 2343 & 3258 & 95.61 & 84.56\\
     CAMixer \cite{CAMixer} & 991 & 2126 & 3117 & 95.80 & 85.35\\
     \rowcolor{Bg} Proposed WBANet & \textbf{605} & 1905 & \textbf{2510} & \textbf{96.62} & \textbf{88.15}\\
    \bottomrule\hline
    \end{tabular}
\end{table}

\subsection{Experimental Results and Discussion}

We evaluated our WBANet against five state-of-the-art methods: CWNN \cite{cwnngao}, SAFNet \cite{SAFNet}, DDNet \cite{DDNet}, LANTNet \cite{LANTNet}, and CAMixer \cite{CAMixer}, implemented using default parameters from their studies. All the experiments, except for the CWNN running with Matlab, were conducted on the Google Colab platform with Python 3.10.12, PyTorch 2.1.0, and an NVIDIA Tesla T4 GPU with 15 GB of memory.

Quantitative results are detailed in Table \ref{table_res}. For the Chao Lake dataset, our WBANet excels in all metrics except false negatives (FN), with significant improvements in the Kappa Coefficient (KC) by 24.37\%, 3.83\%, 2.34\%, 1.18\%, and 0.82\% over CWNN, SAFNet, DDNet, LANTNet, and CAMixer, respectively.
On the Sulzberger dataset, WBANet outperforms other methods in overall error (OE), PCC, and KC. Although CAMixer and CWNN show lower FP and FN rates respectively, they both register higher OEs compared to our WBANet. Similarly, on the Yellow River dataset, WBANet leads in all metrics apart from FN. Notably, it enhances the KC value by 5.69\%, 4.46\%, 4.39\%, 2.79\%, and 2.35\% over CWNN, SAFNet, DDNet, LANTNet, and CAMixer, respectively. While CWNN records a lower FN rate, it lags significantly behind in FP and OE.

Fig. \ref{fig_result} illustrates the visual comparison of change maps produced by different methods on three datasets. Compared to the baseline methods, such as CWNN and SAFNet, our WBANet generates change maps that are visually closer to the ground truth and contain less noise. For instance, in the Yellow River dataset, where speckle noise greatly impacts performance, it is challenging to produce accurate change maps. Here, the performance of CWNN and SAFNet is notably degraded, while DDNet, LANTNet, and CAMixer frequently misclassify changed pixels as unchanged. 

Experimental results on three SAR datasets confirm that our proposed WBANet outperforms other state-of-the-art methods. The effectiveness of our WSM and BAM demonstrates significant contributions to attention feature extraction and non-linear representation modeling. 

\begin{table}[tbp]
\centering
\caption{Ablation studies of the proposed WBANet.}
\setlength{\tabcolsep}{1.5mm}{
\label{table_ablation}
\begin{tabular}{c|c c c} 
\hline\toprule
\multirow{2}{*}{Method} 
    & \multicolumn{3}{c}{PCC on different datasets ($\%$)} \\ \cmidrule{2-4}
& Chao Lake & Sulzberger & Yellow River \\ 
\midrule
Basic Network & 97.46 & 96.02 & 95.67    \\
w/o WSM & 98.13 & 96.33 & 95.92      \\  
w/o BAM & 97.82 & 96.51 & 96.17     \\  
\rowcolor{Bg} Proposed WBANet & 98.33 & 96.65 & 96.62 \\
\bottomrule\hline
\end{tabular}}
\end{table}

\subsection{Ablation Study}

To evaluate the effectiveness of the proposed Wavelet-based Self-attention Block and Bi-dimensional Aggregation Module, ablation experiments were performed on three datasets. We designed three variants: (1) \textit{Basic Network}, which is the WBANet without the WSM and BAM; (2) \textit{w/o WSM}, which omits the Wavelet-based Self-attention Module; and (3) \textit{w/o BAM}, which lacks the Bi-dimensional Aggregation Module. The results in Table~\ref{table_ablation} clearly show that both the WSM and the BAM significantly enhance the non-linear representation capabilities, thereby improving change detection performance.

\begin{figure}[htbp]
  \centering
  \includegraphics[width= 3.35in]{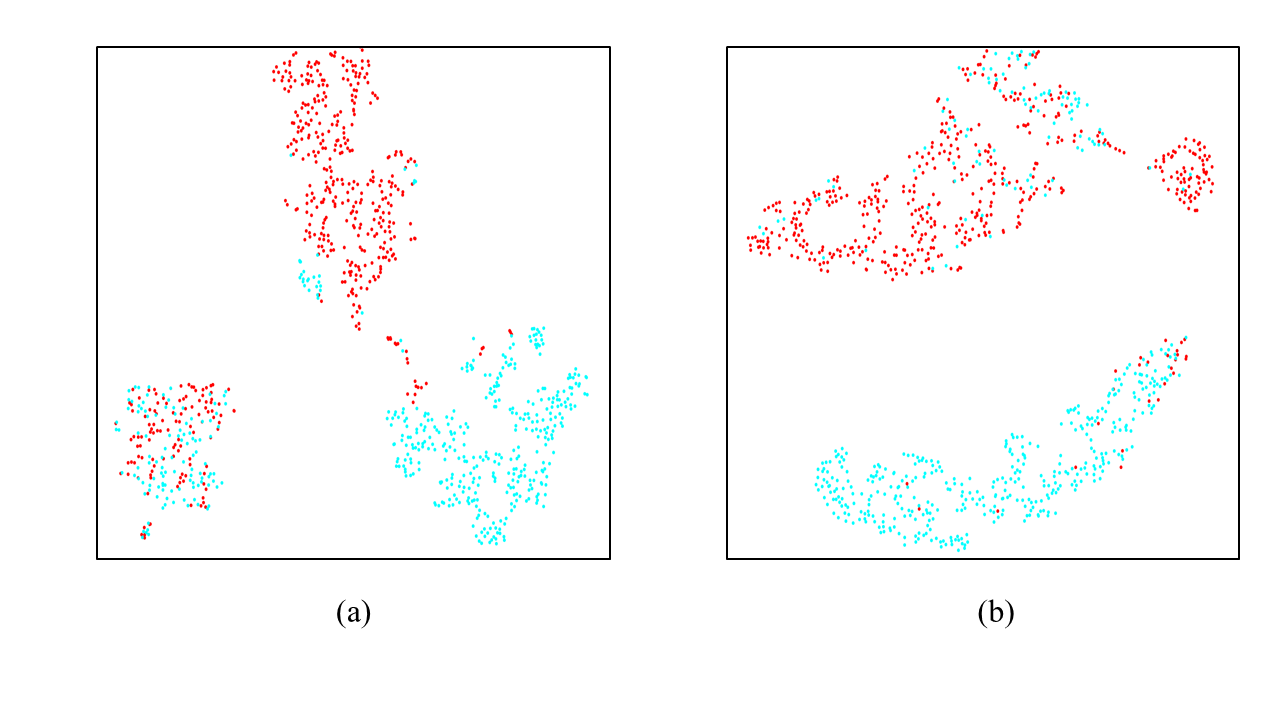}
  \setlength{\abovecaptionskip}{-0.3cm}
  \caption{Visualization of the feature representations on the Yellow River dataset. (a) Features before the WSM. (b) Features after the WSM.}
  \label{fig_visual}
\end{figure}

Additionally, we utilized the t-SNE \cite{tsne} tool to visualize feature characteristics before and after applying the Wavelet-based Self-attention Block. As depicted in Fig. \ref{fig_visual}, the representations post-application display more distinct, well-defined clusters compared to the original input.

\subsection{Analysis of the Block Number}
\begin{figure}[tbp]
  \centering
  \includegraphics[width=3.2in]{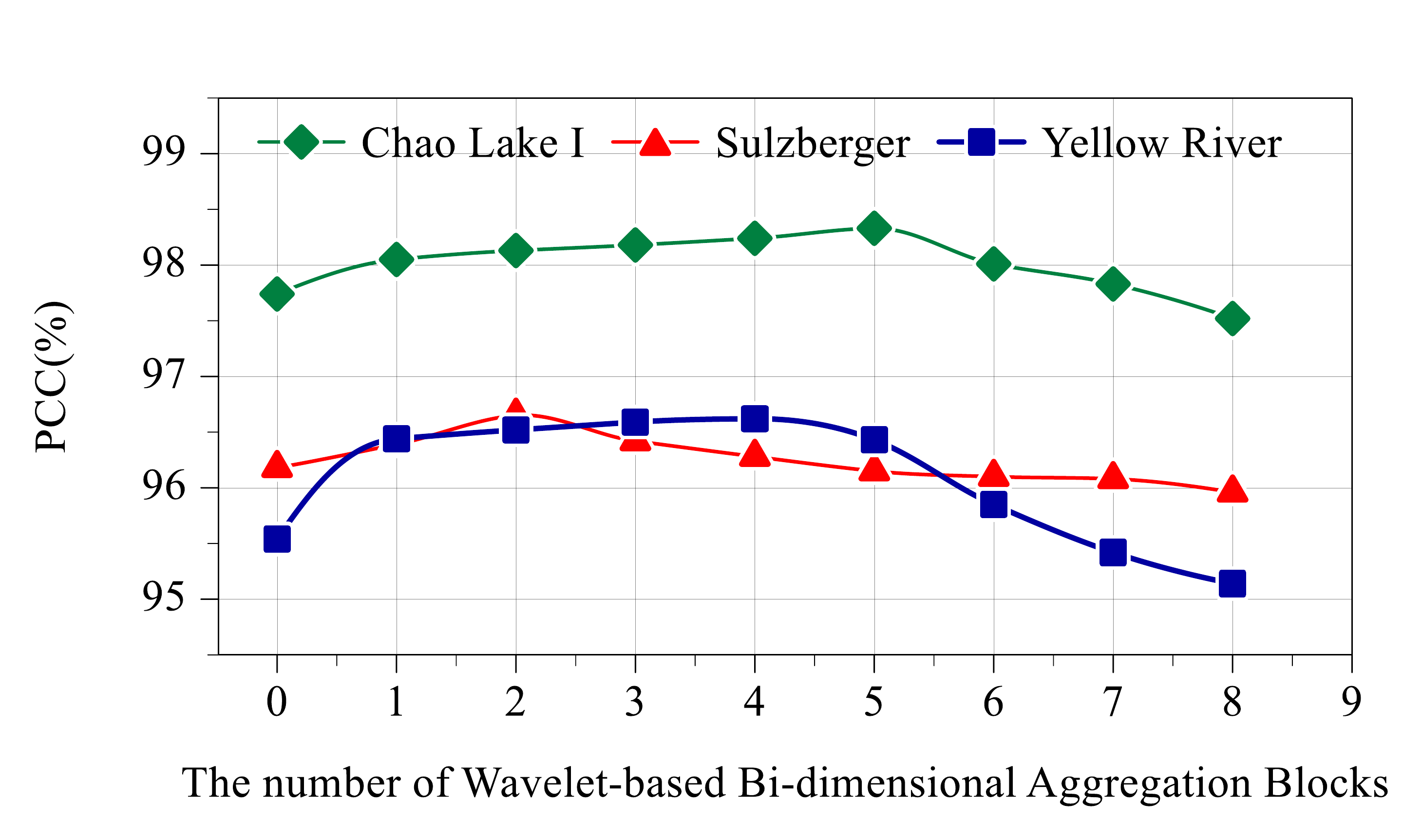}
  \caption{Relationship between the number of attention blocks and PCC values.}
  \label{fig_patchsize}
\end{figure}

The number of Wavelet-based Bi-dimensional Aggregation Blocks, denoted as $N$, is a crucial parameter. We explored the relationship between $N$ and the Percentage of Correct Classification (PCC) by varying $N$ from 0 to 8. As illustrated in Fig. \ref{fig_patchsize}, PCC consistently improves with an increase in the number of Wavelet-based Bi-dimensional Aggregation Blocks up to 5. However, beyond this point, PCC begins to decline due to the increased model complexity. Consequently, we optimized $N$ for different datasets: $N=5$ for the Chao Lake dataset, $N=2$ for the Sulzberger dataset, and $N=4$ for the Yellow River dataset.

\section{Conclusion}
In this letter, we introduce a novel WBANet for SAR image change detection task. The WBANet utilizes DWT and IDWT to achieve down-sampling without the loss of high-frequency details and other important information. Additionally, we developed BAM to enhance non-linear representation capabilities by capturing spatial and channel dependencies and refining features. Extensive experiments on three SAR datasets have verified the effectiveness and rationality of our solution.

\bibliography{reference} 
\bibliographystyle{IEEEtran}

\end{document}